\documentclass[12pt]{article}
\usepackage{amsmath,epsfig,euscript,dsfont}

\def\nslash{\rlap{\hspace{0.02cm}/}{n}}
\def\nbslash{\rlap{\hspace{0.02cm}/}{\bar n}}
\def\kslash{\rlap{\hspace{-0.01cm}/}{k}}

\begin{document}

\begin{titlepage}

\begin{flushright}
EFI 10-14\\
June 16, 2010
\end{flushright}

\vspace{0.7cm}
\begin{center}
\Large\bf\boldmath
An Effective Field Theory Look\\ at\\ Deep Inelastic Scattering
\unboldmath
\end{center}

\vspace{0.8cm}
\begin{center}
{\sc  Gil Paz}\\
\vspace{0.4cm}
{Enrico Fermi Institute\\
The University of Chicago, Chicago, Illinois, 60637, USA
}
\end{center}
\vspace{1.0cm}
\begin{abstract}
  \vspace{0.2cm}
  \noindent
This talk discusses the effective field theory view of deep inelastic
scattering. In such an approach, the standard factorization formula of
a hard coefficient multiplied by a parton distribution function arises
from matching of QCD onto an effective field theory. The DGLAP
equations can then be viewed as the standard renormalization group
equations that determines the cut-off dependence of the non-local
operator whose forward matrix element is the parton distribution
function. As an example, the non-singlet quark splitting functions is
derived directly from the renormalization properties of the non-local
operator itself. This approach, although discussed in the literature,
does not appear to be well known to the larger high energy community.
In this talk we give a pedagogical introduction to this subject.

\end{abstract}
\vfil

\end{titlepage}

\section{Introduction}
Electron-nucleon Deep Inelastic Scattering (DIS) has played an
important role in our understanding of the strong interactions. By
now, it is a standard topic in particle physics and quantum field
theory books \cite{Sterman:1994ce, Peskin:1995ev}.

The cross section for this process can be written in term of two
structure functions. These obey a factorization formula (see for
example \cite{Sterman:1994ce, Sterman:2004pd})
\begin{equation}
F_2(x, Q^2)=\sum_i\, \int\, d\xi\,C_2^i\left(\frac{x}{\xi},\frac{Q}{\mu},\frac{\mu_F}{\mu},\alpha_s(\mu)\right)\,
\phi_{i/h}\Big(\xi,\mu_F,\alpha_s(\mu)\Big)+{\cal O}\left(\frac{\Lambda_{\rm QCD}^2}{Q^2}\right).
\end{equation}

The parton distribution function (PDF) $\phi_{i/h}$ is often referred
to as the ``probability to find a parton $i$ in hadron $h$". Such a
description, although very intuitive, requires a field theoretic
definition. What do we mean by a ``distribution function"?

This question was answered in the early 1980's \cite{Curci:1980uw,
Collins:1981uw}. The quark PDF is
\begin{equation}\label{pdfdef}
\phi_{i/h}(x)=\frac1{2\pi}\int_{-\infty}^{\infty}\,dt\,e^{ix n\cdot p
  \,t}\langle
h(p)|\bar{\psi}_i(0)\,W_n(0)W^\dagger_n(tn)\,\frac{\nslash}{2}\,\psi_i(tn)|h(p)\rangle\Big|_{\rm avg.},
\end{equation}
where $n$ light-like vector, $n^2=0$ and $W_n$ is a light-like Wilson
line,
\begin{equation}\label{Wilson}
 W_n(u)=P\,\exp\,ig\int^0_{-\infty}\,ds\,n\cdot A(u+sn),
\end{equation}
and $A_\mu\equiv A^a_\mu\,t^a$, with $t^a$ in the fundamental
representation. 

Let us discuss (\ref{pdfdef}) in more detail. First, in the definition
of the PDF, an average over color and spin is understood. Second,
since $\bar\psi_i$ and $\psi_i$ are not at the same point, we must
include a ``gauge-link'' to render the operator gauge invariant. This
is the reason that the Wilson lines appear in the definition of the
PDF. Notice that the product of the two semi-infinite Wilson lines can
be written as finite Wilson line connecting the point $0$ and $tn$.
Third, the PDF is just the Fourier transform of the diagonal matrix
element of a standard dimension-three operator. Since the operator is
non-local, its Fourier transform is a function and not a number.
Fourth, the anti-quark PDF and gluon PDF can be defined in a similar
way to the quark PDF. For example, the gluon PDF is defined with the
quark field replaced by gluon field strength and the Wilson line 
is in the adjoint representation. For concreteness
in the following we will focus on the quark PDF.

The PDF satisfies the DGLAP (Dokshitzer-Gribov-Lipatov-Altarelli-Parisi) equation
\begin{equation}
\frac{d\phi_{i/h}(x)}{d\ln \mu}=\int_x^1\frac{d\xi}{\xi}\,P_{ij}(\xi,\alpha_s)\,\phi_{j/h}(x),
\end{equation}
where $P_{ij}$ are the so called ``splitting functions". These are usually 
calculated via two main methods. 

The first is based on the interpretation of the splitting function as
the ``probability'' of the quark to ``split'' into a quark and a gluon
\cite{Altarelli:1977zs}. At ${\cal O}(\alpha_s)$ one finds
$$``P_{qq}"=\frac{C_F\alpha_s}{\pi}\,\frac{1+\xi^2}{1-\xi},$$
where $C_F=4/3$ for QCD.
The full expression is in fact 
\begin{equation}\label{Pqq}
P_{qq}=\frac{C_F\alpha_s}{\pi}\left[(1+\xi^2)\,\frac{1}{\left(1-\xi\right)_+}+\frac32\,\delta(1-\xi)\right]
+{\cal O}(\alpha^2_s)\equiv \frac{C_F\alpha_s}{\pi}P^{[0]}_{qq}+{\cal O}(\alpha^2_s).
\end{equation}
and the singular terms $\dfrac{1}{(1-\xi)_+}$, defined in the appendix, and $\delta(1-\xi)$ must
be added ``by hand".  

The second method, often called the ``OPE'' method
\cite{Georgi:1951sr,Gross:1974cs}, is based on the fact that moments
of the PDF are matrix elements of local operators. Thus one defines
the moments
$$\int_0^1dx\, x^{n-1} \phi(x)=\phi_n$$ 
which satisfy a local renormalization group equation (RGE)
$$\frac{d\phi_n}{d\ln\mu}=-\gamma_n\phi_n.$$
The splitting function is related to $\gamma_n$ via 
$$\gamma_n=-\int_0^1\,dx\,x^{n-1} P_{qq}. $$

There is also another approach possible, which is less known, and 
can be described as the ``effective field theory'' method. Going back
to the factorization formula for $F_2$, we can write it schematically
as
\begin{equation}
\begin{tabular}{ccccccc}\\
$F_2$&=& $C_2(\mu)$ & $\otimes$& $ \phi(\mu)$& +&power corrections\\
&&$\uparrow$&&$\uparrow$&&\\
&&Wilson coef.&&Operator\\
\end{tabular}
\end{equation}
The factorization formula can be interpreted as a result of matching
QCD onto an effective field theory. $C_2$ is the Wilson coefficient
extracted in the process, and the PDF is the matrix element of the
operator in the effective theory. The ``power corrections'' correspond
to the contribution of power suppressed operators. This is the
standard expression one finds in a generic effective field theory
approach. The only complication in this case is the non-locality of
the operator. The scale $\mu$ on which the PDF depends is then just
the cut-off scale of the effective theory. It is not surprising then
that the cut-off dependence of the PDF is determined by an RGE which
is schematically

$$\dfrac{d\phi}{d\ln \mu}=P\otimes\phi.$$

This is none other than the famous DGLAP equation, where the splitting
function is simply the anomalous dimension. Since the PDF is a
function, the anomalous dimension is a function too, and not a
constant.

If the splitting function is just the anomalous dimension of the PDF,
one can calculate it as one usually does in field theory. In
particular, if we use dimensional regularization and a mass
independent scheme, the anomalous dimension can be extracted from the
$1/\epsilon$ pole of the loop corrections to the PDF. The only unusual
aspect of the calculation arises from the non-local structure of the
PDF. We will discuss this calculation in detail in the next section.

This calculation was first performed, to the best of our knowledge, by
Braunschweig, Horejsi and Robaschik \cite{Braunschweig:1982mi}. But
their paper is almost unknown. By the end of 2009 it had 5 citations
on SPIRES \cite{SPIRES}. For comparison, by the end of 2009 reference
\cite{Altarelli:1977zs} had 3933 citations while references
\cite{Georgi:1951sr} and \cite{Gross:1974cs} had 589 and 919
citations, respectively. The situation is slightly better then these
citation counts indicate, but still the third approach is much less
known, especially to the general high energy community.

The effective field theory approach has received much interest in the
recent years, since the effective field theory one is matching onto is
the soft collinear effective theory (SCET) \cite{Bauer:2000yr,
Bauer:2001yt, Beneke:2002ph}.  Calculations of the anomalous dimension of
non-local operators are standard in SCET.
The application of SCET to DIS, as well as other hard QCD processes,
was first discussed in \cite{Bauer:2002nz}. In that paper the matching
coefficients were calculated at tree level. Although SCET is the
appropriate EFT for DIS, the effect of the soft gluons completely
cancel for generic $x$ \cite{Bauer:2002nz, Becher:2006mr}. In this
case SCET is just ``boosted QCD'' and we can calculate the anomalous
dimension by using QCD Feynman rules instead of SCET Feynman rules. We
will demonstrate this approach by calculating $P_{qq}$ for the
non-singlet case. The method will be technically different from
\cite{Braunschweig:1982mi}, but conceptually very similar. For another
pedagogical discussion of the calculation of the evolution (RGE)
equations for non-local operators see appendix G of
\cite{Belitsky:2005qn}, where the calculation techniques are
demonstrated for both covariant and light-cone gauges.

\section{Example of a Non-Local Renormalization:\\
Non-singlet Splitting Function}

\subsection{Feynman rules}
We begin by deriving the Feynman rules. We will use the definition of
the PDF in equation (\ref{pdfdef}), and replace the hadron state by a
free quark. Since we are interested in the cut-off (UV) dependence of
the PDF which is independent of the IR physics, such a replacement is
justified. The ``zero-gluon'' Feynman rule is

\begin{eqnarray}
\vcenter{\epsfig{file=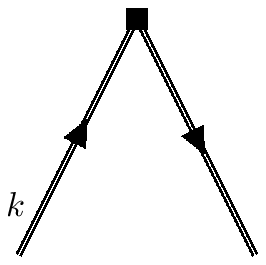,width=1.2cm}}\hspace{-33em}&=&\frac1{2\pi}
\int_{-\infty}^{\infty}\,dt\,e^{i\xi n\cdot p t}\langle
k|\bar{\psi}(0)\,\frac{\nslash}{2}\,\psi(tn)|k\rangle\nonumber\\ 
&=&\frac1{2\pi}\int_{-\infty}^{\infty}\,dt\,e^{i\xi  n\cdot p t}\,
e^{i 0\cdot k}\,e^{-itn\cdot k}\,\bar{u}(k)\,\frac{\nslash}{2}\,u(k)\nonumber\\ 
&=&\delta(\xi\,n\cdot p-n\cdot k)\,\bar{u}(k)\,\frac{\nslash}{2}\,u(k).
\nonumber
\end{eqnarray}
Taking the external states to carry momentum $p$ we find that the
zeroth order expression is
\begin{equation}
D_0=\frac1{n\cdot p}\delta(1-\xi)\bar{u}(p)\,\frac{\nslash}{2}\,u(p).
\end{equation}
Averaging over the spins we would find $\delta(\xi-1)$, i.e. the quark
is carrying all of the momentum of the hadron.  The zero-gluon Feynman
rule is therefore
$$\vcenter{\epsfig{file=uzero.eps,width=1.5cm}} \hspace{-33em}\delta(\xi\,
n\cdot p-n\cdot k)\,\frac{\nslash}{2}.$$

In order to find the one-gluon Feynman rule we need to expand the
Wilson lines in (\ref{Wilson}). A simple calculation yields
$$\vcenter{\epsfig{file=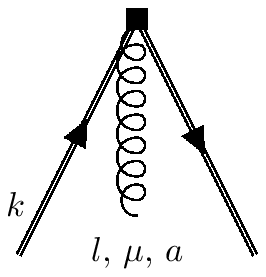,width=1.5cm}} \hspace{-33em}t^a\,g\,\frac{n^\mu}{n\cdot l}
\big[\delta(\xi\, n\cdot p-n\cdot k)-\delta(\xi\, n\cdot p-n\cdot k+n\cdot l)\big]\,\frac{\nslash}{2}$$
\subsection{Feynman integration}
\begin{figure}
\begin{center}
\begin{tabular}{ccc}
\epsfig{file=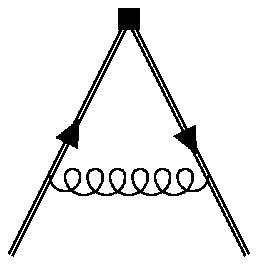,width=2cm}& \epsfig{file=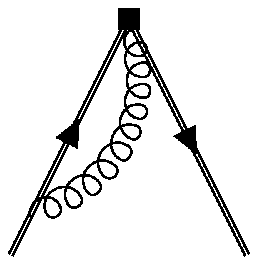,width=2cm}&
\epsfig{file=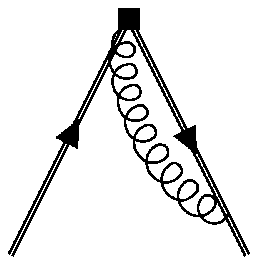,width=2cm}
\end{tabular}
\caption{\label{FD} Feynman diagrams}
\end{center}
\end{figure}

In order to find $P_{qq}$ we need to calculate three diagrams, see
Figure \ref{FD}. To demonstrate the techniques, we will calculate the
leftmost diagram in Figure \ref{FD}. We will use Feynman gauge in the
calculation.

First we need to define light-cone coordinates. We define two
light-cone vectors $n^2=\bar{n}^2=0,\, n\cdot \bar{n}=2$. In
particular we will use $n=(1,0,0,-1),\,\bar{n}=(1,0,0,1)$. Notice that
$n\cdot p= p_0+p_3\geq0$. Any four vector can be decomposed as
\begin{equation}
a^\mu=\bar{n}\cdot a\frac{n^\mu}{2}+n\cdot a\frac{\bar{n}^\mu}{2}+a_\perp^\mu,
\end{equation}
which also defines the ``$\perp$'' coordinates. In these coordinates
$d^dk$ can be written as
$$d^dk=\frac12\,dn\cdot k\,\,d\bar{n}\cdot k\,\,d^{d-2}k_\perp. $$ We
can now write down the Feynman integral, 
\begin{equation}
I=\int\,\frac1{2(2\pi)^d}\,dn\cdot k\,\,d\bar{n}\cdot k\,\,d^{d-2}k_\perp\,\left[\frac{1}{k^2+i0}\right]^2\frac{\rm Num.}{(k-p)^2+i0}
\,\delta(\xi\, n\cdot p-n\cdot k),
\end{equation}
where $p$ ($k$) is the external (internal) quark momentum and
$${\rm Num.}= (n\cdot k)^2\,\frac{\nbslash}2+n\cdot k\,\kslash_\perp-k_\perp^2\,\frac{\nslash}2.$$

The total diagram is given by
\begin{equation}
D_1=-ig^2 C_F \mu^{2\epsilon}\, \bar u(p)\,\gamma^\mu I\gamma_\mu u(p)
\end{equation}

The only difference from the usual Feynman integral is the presence
of the delta function. It also forces us to split the integration
according to the light-cone coordinates. Our strategy will be to
calculate the $\bar n\cdot k$ integral using residue theorem, to
calculate the $k_\perp$ integral in $d-2$ dimensions, and use the
delta function to perform the integral over $n\cdot k$
\cite{Grozin:1996pq}.

In order to use the residue theorem for the $\bar n\cdot k$ integral
we need to find the poles of the integrand. They are given by
\begin{eqnarray}\label{poles}
&&k^2+i0=\bar{n}\cdot k\,n\cdot k+k_\perp^2+i0=0\quad\Rightarrow\quad\bar{n}\cdot k=\dfrac{-k_\perp^2-i0}{n\cdot k}
\nonumber\\
&&(k-p)^2+i0=0\quad\Rightarrow\quad\bar{n}\cdot k=\bar n\cdot p+\dfrac{-(k-p)_\perp^2-i0}{n\cdot k-n\cdot p}.
\end{eqnarray}
In order to get a non-zero result we must ensure that the two poles
are on the opposite sides of the real $\bar n\cdot k$ axis. Otherwise
we can close the contour in the half plane that contains no poles and
get zero. There are two options then,
\begin{equation}
\begin{tabular}{cccccccc}
I)&$n\cdot k>0$&$\cup$&$n\cdot k-n\cdot p<0$&$\Rightarrow$&$0<n\cdot k<n\cdot p$&$\Rightarrow$&$0<\dfrac{n\cdot k}{n\cdot p}<1$\\
II)&$n\cdot k<0$&$\cup$&$n\cdot k-n\cdot
p>0$&$\Rightarrow$&$n\cdot p<n\cdot k<0$.&&\\
\end{tabular}
\end{equation}
Recall, though, that $n\cdot p>0$, so the second option is ruled
out. Combining the first option with the delta function $\delta(\xi\,
n\cdot p-n\cdot k)$ we find that $0<\xi<1$. Notice also that we have
recovered the usual support property of $\phi_{i/h}(\xi)$, namely, that
the momentum fraction carried by the quark must be between 0 and 1.

After performing the $n\cdot k$ integral using residues, we perform
the integral over $k_\perp$ using dimensional regularization in $d-2$
dimensions. Due to the numerator structure, we can split $I$ into 3
integrals that differ by their Dirac structure. In order to find the
anomalous dimension we only need the Dirac structure that appear in
the definition of $\phi_{i/h}(\xi)$, i.e $\nslash$. The other Dirac
structures lead to finite terms and do not contribute to the anomalous
dimension.
\footnote{We can verify this point using another argument. The integral can be
organized by the so called method of regions
\cite{Beneke:1997zp,Smirnov:2002pj}, which is also related to our
effective field theory interpretation. The method of regions allows us
to divide the momentum integration to various regions in which the
integral momentum scales in a particular way. The sum of all the
regions is equal to the full integral.

For our integrals there is only a collinear region, namely a region in
which the components of the momentum $k$ scale as

$$n\cdot k\sim{\cal{O}}(1),\quad \bar n\cdot k\sim{\cal{O}}(\Lambda^2_{\rm QCD}),\quad k_\perp\sim {\cal{O}}(\Lambda_{\rm QCD}).$$

We also have 

$$d^4k\sim{\cal{O}}(\Lambda^4_{\rm QCD}),\quad k^2\sim
{\cal{O}}(\Lambda^2_{\rm QCD})\quad \Rightarrow I\sim \Lambda^4_{\rm
  QCD}\,\left(\frac1{\Lambda^2_{\rm QCD}}\right)^3\,\times{\rm Num.}$$

Since the anomalous dimension can depends only on $n\cdot p$, it must
be an ${\cal{O}}(1)$ quantity, and so does $I$. We therefore need the numerator to
scale as $\Lambda^2_{\rm QCD}$, so only the $k_\perp^2$ integral will
contribute to the anomalous dimension. This integral involves only the
$\nslash$ Dirac structure.} 

We will therefore ignore the other Dirac
structures in the numerator. In order to regulate the IR divergences we keep $p^2\neq 0$. 
Performing the $\bar n\cdot k$ integral for this structure by using
the first pole in (\ref{poles}), we find
\begin{eqnarray}
I_{\nslash}&=&-\frac12\int\,\frac1{2(2\pi)^d}\,dn\cdot k\,\,d\bar{n}\cdot k\,\,d^{d-2}k_\perp\,\left[\frac{1}{k^2+i0}\right]^2\frac{k^2_\perp \nslash}{(k-p)^2+i0}
\,\delta(\xi\, n\cdot p-n\cdot k)\nonumber\\
&=&i\frac{\nslash}{2} \frac1{4\pi}\int_0^1 dr\, \delta(\xi-r)\int\,\frac{d^{d-2}k_\perp}{(2\pi)^{d-2}}
\,\frac1{n\cdot p}\,\frac{(1-r)(k_\perp+rp_\perp)^2}{[k^2_\perp-(-p^2-i0) r(1-r)]^2}
\end{eqnarray}
where we have defined $r=n\cdot k/n\cdot p$. The limits of integration
are determined by the $\bar n \cdot k$ integral as explained
above. Performing the relevant $k_\perp$ integration, we find that the
divergent part of the diagram is
\begin{eqnarray}
D_1^{\rm div}&=&-\frac{1}{n \cdot p}\frac{C_F\alpha_s}{4\pi}\, \bar u(p)\,\gamma^\mu \frac{\nslash}{2}\gamma_\mu u(p)\,(4\pi\mu^2)^\epsilon(1-\epsilon)\Gamma(\epsilon)(-p^2-i0)^{-\epsilon}\nonumber\\
&\times&\int_0^1 dr\,  (1-r)[r(1-r)]^{-\epsilon}\delta(\xi-r),
\end{eqnarray}
 Expanding in $\epsilon$ we finally find
\begin{equation}
D_1^{\rm div}=\frac1{\epsilon}\frac{1}{n \cdot p}\frac{C_F\alpha_s}{4\pi}\, \bar u(p)\,\frac{\nslash}{2} u(p)\,
 2 (1-\xi) \theta (\xi)\,\theta(1-\xi).
\end{equation}
The other two diagrams can be calculated using the same methods. Using the one-gluon Feynman rule we find that
\begin{eqnarray}
D_2=D_3&=&\frac{1}{n \cdot p}\frac{C_F\alpha_s}{4\pi}\, 
\bar u(p)\,\frac{\nslash}{2} u(p)\,(4\pi\mu^2)^\epsilon\,\Gamma(\epsilon)(-p^2-i0)^{-\epsilon}\nonumber\\
&\times&\Big[\xi^{1-\epsilon}(1-\xi)^{-1-\epsilon}\theta(1-\xi)\theta(\xi)-\delta(1-\xi)\,{\rm B}(2-\epsilon,-\epsilon)\Big].
\end{eqnarray}
In order to expand in $\epsilon$ we need to use the identity \cite{Gelfand:1964}, which is also proven in the appendix,
\begin{equation}
\theta(\xi)\theta(1-\xi)(1-\xi)^{-1-\epsilon}=-\frac1\epsilon\,\delta(1-\xi)+\left(\frac1{1-\xi}\right)_++{\cal O}(\epsilon). 
\end{equation}
Expanding in $\epsilon$ we finally find that the divergent part is
\begin{eqnarray}
D_2^{\rm div}=D_3^{\rm div}&=&\frac1{\epsilon}\,\frac{1}{n \cdot p}\frac{C_F\alpha_s}{4\pi}\, 
\bar u(p)\,\frac{\nslash}{2} u(p)\,2\,\theta(\xi)\Big[
\xi\,\frac{1}{(1-\xi)_+}+\delta(1-\xi)\Big].
\end{eqnarray}
Notice that the singular terms $\dfrac{1}{(1-\xi)_+}$ and
$\delta(1-\xi)$ arise naturally in the course of the calculation and are
not added ``by hand".

We also need the wave function renormalization constant $Z_Q$ \cite{Grozin:2005yg} in Feynman gauge
\begin{equation}
Z_Q=1-\frac{C_F\alpha_s}{4\pi}\frac1\epsilon.
\end{equation}
The total expression for the matrix element at one loop accuracy is 
\begin{eqnarray}
D^{\rm div}_{\rm total}&=&D^{\rm div}_1+D^{\rm div}_2+D^{\rm
div}_3+Z_Q\,D_0\nonumber\\ &=&\frac1{n\cdot
p}\bar{u}(p)\,\frac{\nslash}{2}\,u(p)\bigg\{\delta(1-\xi)+\theta(\xi)\,\theta(1-\xi)\,\frac{C_F\alpha_s}{4\pi}\frac2\epsilon\Big[(1-\xi)+2\xi\,\frac{1}{(1-\xi)_+}+\frac32\delta(1-\xi)\Big]\bigg\}\nonumber\\
&=&\frac1{n\cdot
p}\bar{u}(p)\,\frac{\nslash}{2}\,u(p)\bigg\{\delta(1-\xi)+\theta(\xi)\,\theta(1-\xi)\,\frac{C_F\alpha_s}{4\pi}\frac2\epsilon
\Big[\left(1+\xi^2\right)\frac{1}{(1-\xi)_+}+\frac32\,\delta(1-\xi)\Big]\bigg\}.\nonumber\\
\end{eqnarray}
Averaging over the spins we find that
\begin{eqnarray}
\phi_{\rm bare}(\xi)&=&\delta(1-\xi)+\frac{C_F\alpha_s}{4\pi}\,
\frac2\epsilon\,\theta(\xi)\,\theta(1-\xi)\,\Big[\left(1+\xi^2\right)\frac{1}{(1-\xi)_+}+\frac32\,\delta(1-\xi)\Big]+\rm finite+{\cal O}(\alpha^2_s)\nonumber\\
\end{eqnarray}
\subsection{Renormalization}
We now calculate the RGE equation in the $\overline{\rm MS}$ scheme in
the ``standard way'', see for example \cite{Buras:1998raa}. We define
the renormalization factor $Z(x,\xi)$ by
\begin{equation}
\phi^{\rm bare}(x)=\int_0^1 d\xi\, Z(x,\xi)\,\phi^{\rm ren.}(\xi)\equiv Z(x,\xi)\otimes\phi^{\rm ren.}(x) ,
\end{equation}
and expand $Z$ in $\alpha_s/4\pi$, such that $Z_{[n]}$ denotes the
coefficient of the $(\alpha_s/4\pi)^n$ term. At tree level we have
$\phi_{[0]}^{\rm bare}(x)=\phi_{[0]}^{\rm ren.}(x)=\delta(1-x)$, so
\begin{equation}
Z_{[0]}(x,\xi)=\delta(\xi-x)=\frac1\xi\,\delta(1-\frac{x}{\xi}).
\end{equation} 
In order to calculate the one loop expression for the renormalization
factor we use $$\phi_{[1]}^{\rm
bare}(x)=Z_{[0]}(x,\xi)\otimes\phi_{[1]}^{\rm
ren.}(\xi)+Z_{[1]}(x,\xi)\otimes\phi_{[0]}^{\rm ren.}(\xi)$$ to find
\begin{equation}
Z_{[1]}(x,\xi)=\theta(1-\frac{x}{\xi})\theta(\frac{x}{\xi})\,\frac1\epsilon\,\frac2{\xi}\,P^{[0]}_{qq}\left(\frac{x}{\xi}\right),
\end{equation}
or
\begin{equation}
Z(x,\xi)=\frac1\xi\,\delta(1-\frac{x}{\xi})+\frac{C_F\alpha_s}{4\pi}\theta(1-\frac{x}{\xi})
\theta(\frac{x}{\xi})\,\frac1\epsilon\,\frac2{\xi}\,P^{[0]}_{qq}\left(\frac{x}{\xi}\right)+{\cal
O}(\alpha^2_s),
\end{equation}
where $P_{qq}^{[0]}$ is defined in (\ref{Pqq}). In order to calculate
the anomalous dimension we now follow the standard procedure
\cite{Buras:1998raa}, treating $Z$ and $\gamma$ as infinite
dimensional matrices. Define
\begin{eqnarray}
O^{\rm bare}=Z\otimes O^{\rm ren.}\quad{\rm and} \quad 
\frac{dO^{\rm ren.}}{d\ln\mu}=\gamma\otimes O^{\rm ren.},
\end{eqnarray}
and expand $Z$ as
\begin{equation}
Z=\mathds{1}+\sum_{k=1}^\infty\frac1{\epsilon^k}Z^{(k)}.
\end{equation}
The anomalous dimension $\gamma$ is given by
\begin{equation}
\gamma=2\alpha_s\frac{\partial Z^{(1)}}{\partial\alpha_s}.
\end{equation}
In this way we find that 
\begin{equation}
\gamma(x,\xi)=\frac{C_F\alpha_s}{\pi}\theta(1-\frac{x}{\xi})\theta(\frac{x}{\xi})\,\frac1{\xi}\,P^{[0]}_{qq}\left(\frac{x}{\xi}\right)+{\cal O}(\alpha^2_s),
\end{equation}
and the RGE is finally
\begin{equation}
\frac{d\phi(x)}{d\ln\mu}=\int_0^1\,d\xi\,\gamma(x,\xi)\,\phi(\xi)=
\int_x^1\,\frac{d\xi}{\xi}\,P^{[0]}_{qq}\left(\xi\right)\phi\left(\frac{x}{\xi}\right)+{\cal O}(\alpha^2_s),
\end{equation}
where we have changed $\xi\to x/\xi$. We have recovered the standard DGLAP equation at ${\cal O}(\alpha_s)$. 
\section{Conclusions}

We have argued that the standard features of DIS, namely,
factorization into a short distance coefficient times a parton
distribution function and DGLAP evolution of the PDF, can be
understood in the language of an effective field theory. In this
interpretation QCD is matched onto an effective field theory. The
short distance coefficient is simply the Wilson coefficient and the
PDF is the matrix element of the operator in the effective field
theory. The only difference from more familiar effective field
theories is that the operator is non-local and therefore its matrix
element is a function and not a number. The cut-off dependence of the
matrix element can be determined in the standard way. In particular,
by using dimensional regularization and a mass independent scheme, the
anomalous dimension can be extracted from the $1/\epsilon$ pole of the
loop corrections to the PDF. The resulting RGE equations are none
other than the famous DGLAP equations. This approach, although
discussed in the literature, does not appear to be well known to the
larger high energy community.

We have demonstrated this approach explicitly by calculating the
one-loop correction to the non-singlet quark PDF. We have then derived
the evolution equation from the divergent piece of the one-loop
correction. This procedure can be generalized in several ways.  First,
we can calculate higher order corrections in $\alpha_s$.  Second, we
can include operator mixing, such as the calculation of $P_{qg}$ and
$P_{gq}$. Third, we can include power corrections. For example, one
can calculate in this method the anomalous dimension of twist 4
operators, which is still an open problem.

\vspace{0.3cm} 
{\em Acknowledgments:} I would like to thank Donal O'Connell for
useful discussions and his comments on the manuscript. I would also
like to thank several of my colleagues who expressed interest in the
topic and encouraged me to write up this talk. This work is supported
in part by the Department of Energy grant DE-FG02-90ER40560.

\begin{appendix}
\section{The plus distribution}
The ``plus" distribution $\dfrac{1}{(1-x)_+}$ is defined such that for any test function $f(x)$
\begin{equation}
\int_0^1 dx\,f(x)\, \frac{1}{(1-x)_+}=\int_0^1 dx\,\frac{f(x)-f(1)}{(1-x)}.
\end{equation}
More generally $\big[h(x)\big]_+$ is defined as 
\begin{equation}
\int_0^1 dx\,f(x)\, \big[h(x)\big]_+=\int_0^1 dx\,\Big(f(x)-f(1)\Big)\,h(x).
\end{equation}
When multiplying  $\big[h(x)\big]_+$ by a function $g(x)$ which is regular at $x=1$ we have
\begin{equation}
\int_0^1 dx\,f(x)\, g(x)\,\big[h(x)\big]_+=\int_0^1 dx\,\Big(f(x)g(x)-f(1)g(1)\Big)h(x).
\end{equation} 
Notice that 
\begin{equation}
\big[g(x)h(x)\big]_+=g(x)\big[h(x)\big]_+-\delta(1-x)\int_0^1 dx\,\Big(g(x)-g(1)\Big)\,h(x).
\end{equation}
Another useful property is the identity
\begin{equation}
(1-x)^n\,\frac{1}{(1-x)_+}=(1-x)^{n-1}.
\end{equation}
Thus we can write
\begin{equation}
P^{[0]}_{qq}=(1-\xi)+2\xi\,\frac{1}{(1-\xi)_+}+\frac32\,\delta(1-\xi)=(1+\xi^2)\,\frac{1}{\big(1-\xi\big)_+}+\frac32\,\delta(1-\xi)=\left(\frac{1+\xi^2}{1-\xi}\right)_+.
\end{equation}
In order to prove the identity
\begin{equation}
\theta(x)\theta(1-x)(1-x)^{-1-\epsilon}=-\frac1\epsilon\,\delta(1-x)+\left(\frac1{1-x}\right)_++{\cal O}(\epsilon), 
\end{equation}
we multiply $(1-x)^{-1-\epsilon}$ by a test function, $f(x)$ and integrate
\begin{equation}
\int_0^1 dx\,f(x)\,(1-x)^{-1-\epsilon}=\int_0^1 dx\,\big[f(x)-f(1)\big]\,(1-x)^{-1-\epsilon}+f(1)\,\int_0^1 dx\,(1-x)^{-1-\epsilon}
\end{equation}
Both terms on the right hand side are regular at $x=1$, so we can expand in $\epsilon$ and find
\begin{equation}
\int_0^1 dx\,f(x)\,(1-x)^{-1-\epsilon}=
\int_0^1 dx\,\frac{f(x)-f(1)}{1-x}-\frac{f(1)}{\epsilon}+{\cal O}(\epsilon).
\end{equation}
Notice that 
\begin{equation}
\theta(x)\theta(1-x)\,g(x)\,(1-x)^{-1-\epsilon}=-\frac1\epsilon\,\delta(1-x)\,g(1)+g(x)\,\left(\frac1{1-x}\right)_++{\cal O}(\epsilon).
\end{equation}

\end{appendix}


\begin{thebibliography}{99}
\bibitem{Sterman:1994ce}
  G.~Sterman,
 ``An Introduction to quantum field theory,''
  {Cambridge, UK: Univ. Pr. (1993) 572 p}

\bibitem{Peskin:1995ev}
  M.~E.~Peskin and D.~V.~Schroeder,
  ``An Introduction To Quantum Field Theory,''
{  Reading, USA: Addison-Wesley (1995) 842 p}

\bibitem{Sterman:2004pd} 
G.~Sterman,
  arXiv:hep-ph/0412013.

\bibitem{Curci:1980uw}
  G.~Curci, W.~Furmanski and R.~Petronzio,
  Nucl.\ Phys.\  B {\bf 175}, 27 (1980).

\bibitem{Collins:1981uw}
  J.~C.~Collins and D.~E.~Soper,
  Nucl.\ Phys.\  B {\bf 194}, 445 (1982).

\bibitem{Altarelli:1977zs}
  G.~Altarelli and G.~Parisi,
  Nucl.\ Phys.\  B {\bf 126}, 298 (1977).

\bibitem{Georgi:1951sr}
  H.~Georgi and H.~D.~Politzer,
  Phys.\ Rev.\  D {\bf 9}, 416 (1974).

\bibitem{Gross:1974cs}
  D.~J.~Gross and F.~Wilczek,
  Phys.\ Rev.\  D {\bf 9}, 980 (1974).

\bibitem{Braunschweig:1982mi}
  T.~Braunschweig, J.~Horejsi and D.~Robaschik,
  Z.\ Phys.\  C {\bf 23}, 19 (1984).

\bibitem{SPIRES}
http://www.slac.stanford.edu/spires/

\bibitem{Bauer:2000yr}
  C.~W.~Bauer, S.~Fleming, D.~Pirjol and I.~W.~Stewart,
  Phys.\ Rev.\  D {\bf 63}, 114020 (2001)
  [arXiv:hep-ph/0011336].

\bibitem{Bauer:2001yt}
  C.~W.~Bauer, D.~Pirjol and I.~W.~Stewart,
  Phys.\ Rev.\  D {\bf 65}, 054022 (2002)
  [arXiv:hep-ph/0109045].

\bibitem{Beneke:2002ph}
  M.~Beneke, A.~P.~Chapovsky, M.~Diehl and T.~Feldmann,
  Nucl.\ Phys.\  B {\bf 643}, 431 (2002)
  [arXiv:hep-ph/0206152].

\bibitem{Bauer:2002nz}
  C.~W.~Bauer, S.~Fleming, D.~Pirjol, I.~Z.~Rothstein and I.~W.~Stewart,
  Phys.\ Rev.\  D {\bf 66}, 014017 (2002)
  [arXiv:hep-ph/0202088].

\bibitem{Becher:2006mr}
  T.~Becher, M.~Neubert and B.~D.~Pecjak,
  JHEP {\bf 0701}, 076 (2007)
  [arXiv:hep-ph/0607228].

\bibitem{Belitsky:2005qn}
  A.~V.~Belitsky and A.~V.~Radyushkin,
  Phys.\ Rept.\  {\bf 418}, 1 (2005)
  [arXiv:hep-ph/0504030].

\bibitem{Grozin:1996pq}
  A.~G.~Grozin and M.~Neubert,
  Phys.\ Rev.\  D {\bf 55}, 272 (1997)
  [arXiv:hep-ph/9607366].


\bibitem{Beneke:1997zp}
  M.~Beneke and V.~A.~Smirnov,
  Nucl.\ Phys.\  B {\bf 522}, 321 (1998)
  [arXiv:hep-ph/9711391].

\bibitem{Smirnov:2002pj}
  V.~A.~Smirnov,
  Springer Tracts Mod.\ Phys.\  {\bf 177}, 1 (2002).

\bibitem{Gelfand:1964}
I.M. Gelfand and G.E. Shilov, 
”Generalized functions”, 
New York, USA: Academic Press (1964)

\bibitem{Grozin:2005yg}
  A.~Grozin,
  arXiv:hep-ph/0508242.

\bibitem{Buras:1998raa}
  A.~J.~Buras,
  arXiv:hep-ph/9806471.







\end{thebibliography}
\end{document}